\documentclass[12pt]{article}

\usepackage{graphicx}
\usepackage{float}
\usepackage{cite}
\usepackage{amsfonts}
\usepackage{amssymb}
\usepackage{amsmath}
\usepackage{relsize}
\usepackage[compatibility=false]{caption}
\usepackage{subcaption}
\usepackage{xcolor}

\def\gtwid{\mathrel{\raise.3ex\hbox{$>$\kern-.75em\lower1ex\hbox{$\sim$}}}}
\def\ltwid{\mathrel{\raise.3ex\hbox{$<$\kern-.75em\lower1ex\hbox{$\sim$}}}}
\def\square{\kern1pt\vbox{\hrule height 1.2pt\hbox{\vrule width 1.2pt\hskip 3pt
   \vbox{\vskip 6pt}\hskip 3pt\vrule width 0.6pt}\hrule height 0.6pt}\kern1pt}

\begin{document}

\begin{titlepage}

\begin{flushright}
UFIFT-QG-25-05
\end{flushright}

\vskip 2.5cm

\begin{center}
{\bf A Nonlocal Realization of MOND that Interpolates from
Cosmology to Gravitationally Bound Systems}
\end{center}

\vskip 1cm

\begin{center}
C. Deffayet$^{1*}$ and R. P. Woodard$^{2\dagger}$
\end{center}

\begin{center}
\it{$^{1}$ Laboratoire de Physique de l'Ecole Normale Sup\'{e}rieure,
ENS, Universit\'{e} PSL, CNRS
Sorbonne Universit\'{e}, Universit\'{e} Paris Cit\'{e}, F-75005 Paris, FRANCE}
\end{center}

\begin{center}
\it{$^{2}$ Department of Physics, University of Florida,\\
Gainesville, FL 32611, UNITED STATES}
\end{center}

\vspace{1cm}

\begin{center}
ABSTRACT
\end{center}
Nonlocal modifications of gravity derive from corrections to the quantum
gravitational stress tensor which grow nonperturbatively strong during
primordial inflation and may persist to the current epoch. Phenomenological 
constructions have been given that realize MOND in gravitationally bound
systems and, separately, reproduce all the cosmological phenomena usually
ascribed to dark matter, including the cosmic microwave background radiation, 
baryon acoustic oscillations and linearized structure formation. In this work 
we exhibit a single model that interpolates between the two regimes.

\begin{flushleft}
PACS numbers: 04.50.Kd, 95.35.+d, 98.62.-g
\end{flushleft}

\vspace{2cm}

\begin{flushleft}
$^{*}$ e-mail: cedric.deffayet@phys.ens.fr \\
$^{\dagger}$ e-mail: woodard@phys.ufl.edu
\end{flushleft}

\end{titlepage}

\section{Introduction}

The motivation for considering nonlocal modifications of gravity derives
from the phenomenon of inflationary particle production, which is responsible
for the power spectra of primordial gravitons \cite{Starobinsky:1979ty} and
scalars \cite{Mukhanov:1981xt}. The occupation number of a single wave vector
$\vec{k}$ of a massless, minimally coupled scalar, or a single polarization 
of the graviton, grows at a staggering rate in de Sitter background,
\begin{equation}
N(t,k) = [ \tfrac{H a(t)}{2 c k}]^2 \qquad , \qquad a(t) = e^{H t} \; .
\label{Ocnumber}
\end{equation}
These quanta alter the kinematics of gravitational radiation and the force
of gravity by factors of $\ln[ a(t)]$ which grow without bound. For example,
loops of massless, minimally coupled scalars change the electric components
of the Weyl tensor for plane wave radiation, and the Newtonian potential
induced by a point mass $M$, to \cite{Park:2015kua,Miao:2024atw},
\begin{eqnarray}
C_{0i0j} &\!\!\! = \!\!\!& C^{\rm tree}_{0i0j} \Bigl\{1 \!-\!
\tfrac{3 \hbar G H^2}{10 \pi c^5} \ln[a] \!+\! \dots\Bigr\} \longrightarrow
C^{\rm tree}_{0i0j} \!\times\! [ a(t)]^{-\frac{3 \hbar G H^2}{10 \pi c^5}} ,
\label{Weyl} \\
\Psi &\!\!\! = \!\!\!& \tfrac{G M}{a r} \Bigl\{\!1 \!+\! \tfrac{\hbar G}{20 \pi
c^3 a^2 r^2} \!-\! \tfrac{3 \hbar G H^2}{10 \pi c^5} \ln[\tfrac{a H r}{c}]
\!+\! \dots\!\Bigr\} \longrightarrow \tfrac{G M}{a r} \!\times\!
[ \tfrac{a(t) H r}{c}]^{-\frac{3 \hbar G H^2}{10 \pi c^5}} . \quad
\label{Newtonscalar}
\end{eqnarray}
The middle results are from explicit, 1-loop computations, whereas the results
on the far right represent nonperturbative resummations of the leading 
logarithms at each order \cite{Miao:2024nsz}.

The fact that quantum corrections to the force of gravity, such as
(\ref{Newtonscalar}), grow nonperturbatively strong during inflation suggests
that the late-time phenomena usually ascribed to dark matter might be due
instead to nonlocal modifications of gravity from the effective action of
quantum gravity. Although the development of a leading logarithm resummation 
for quantum gravity has made impressive recent progress \cite{Miao:2024shs,
Miao:2025gzm,Miao:2025bmd}, it has not yet produced an explicit model.
Therefore, models which might replace dark matter have so far been explored
on a purely phenomenological basis. This amounts to making guesses about 
what might be the macroscopically most significant part of the effective 
action.

The first attempt was based on an algebraic function of the inverse scalar
d'Alembertian acting on the Ricci scalar because that nonlocal invariant
degenerates to $-4 \ln[a(t)]$ when specialized to de Sitter 
\cite{Tsamis:1997rk}. The resulting model produced flat galactic rotation 
curves without dark matter, but failed to explain the observed level of 
weak lensing \cite{Soussa:2003vv}. A more elaborate class of models was 
devised based on the inverse scalar d'Alembertian acting on the Ricci 
tensor contacted into two timelike 4-velocity fields. This class of models
was successful in describing gravitationally bound structures without dark
matter \cite{Deffayet:2011sk,Deffayet:2014lba} but it failed when extended 
to cosmology \cite{Kim:2016nnd,Tan:2018bfp}.

A recent model was constructed to exactly reproduce the cosmological 
successes of dark matter \cite{Deffayet:2024ciu}. The construction
exploits the fact that dark matter behaves, on cosmological scales, like
a perfect fluid with zero pressure and no direct interactions with other
species except gravity \cite{Tulin:2017ara,Sarkar:2018xon,Adhikari:2022sbh}.
This means that the dark matter stress tensor must be separately conserved,
which provides enough equations to determine it as a nonlocal functional of
the metric. 

The purpose of this paper is to exhibit a single model that interpolates
between the cosmological and gravitationally bound regimes, and recovers 
the previous successful models in each case. Section 2 describes the 
cosmological regime, and section 3 describes the gravitationally bound 
regime. In section 4 we show how the cosmological model can be reformulated
and then extended so that it recovers the gravitationally bound model in
regions for which the spatial dependence of the metric dominates over its
time dependence. Our conclusions comprise section 5.

\section{The Cosmological Regime}

The purpose of this section is to review the successful cosmological model
\cite{Deffayet:2024ciu}. We begin by explaining how to construct a timelike
vector field $u_{\mu} = \partial_{\mu} \phi$ as the gradient of a scalar 
which obeys a first order equation. We then show how conservation of the 
dark matter stress tensor $T_{\mu\nu} = \rho u_{\mu} u_{\nu}$ gives a first 
order equation for the CDM energy density $\rho$. In 
demonstrating that the equations for $\phi[g]$ and $\rho[g]$ are well-posed 
we employ the ADM (Arnowitt-Deser-Misner) representation of a general metric 
\cite{Arnowitt:1959ah},
\begin{equation}
ds^2 = -N^2 dt^2 + \gamma_{ij} (dx^i - N^i dt) (dx^j - N^j dt) \; ,
\label{ADM}
\end{equation}
where $N(t,\vec{x})$ is the lapse, $N^i(t,\vec{x})$ is the shift and
$\gamma_{ij}(t,\vec{x})$ is the 3-metric. The section closes by presenting
a Lagrangian whose variation reproduces the equations for $\phi$ and $\rho$, 
as well as the dark matter stress tensor.

\subsection{A Timelike 4-Velocity Field}

The simplest way to construct a timelike vector field $u_{\mu}[g](x)$ from 
the metric is as the gradient of a scalar $\phi[g](x)$ that obeys a first 
order equation with null initial value data \cite{Deffayet:2024ciu},
\begin{equation}
\partial_{\mu} \phi \partial_{\nu} \phi g^{\mu\nu} = -1 \qquad , \qquad
\phi(0,\vec{x}) = 0 \; . \label{phieqn}
\end{equation}
Here $t=0$ is the time, in the distant past, when primordial
inflation ended. Expressing this equation in terms of the ADM variables
(\ref{ADM}) and then solving for the time derivative of the scalar gives,
\begin{equation}
\dot{\phi} = N \sqrt{1 + \gamma^{ij} \partial_i \phi \partial_j \phi} -
N^i \partial_i \phi \; . \label{phieqn2}
\end{equation}
This is obviously a well-posed equation, which determines the scalar
$\phi[g](t,\vec{x})$ as a unique, nonlocal functional of the metric obeying
the initial condition $\phi(0,\vec{x}) = 0$.

A potential pathology of this system would be the formation of
caustics at which the 4-velocity $u_{\mu}(x) = \partial_{\mu} \phi(x)$ 
becomes multi-valued. Because equation (\ref{phieqn2}) seems to produce a
unique, single-valued solution for the scalar, the formation of such a caustic
could only happen if $\phi(t,\vec{x})$ developed a cusp in space. In our 
context, a smooth geometry is enough to yield a smooth evolution in time of 
(\ref{phieqn2}).  The belief that caustics should nonetheless form seems to 
be based in part on an alternate representation of a pressureless, perfect 
fluid as a dense collection of freely falling, point particles 
\cite{Felder:2002sv,Blas:2009yd}. These particles follow geodesics 
$\chi^{\mu}(\tau)$ whose fall into --- and then back out of --- even a 
smooth, central gravitational field must lead to particles being at the same 
point in space with different 4-velocities. However, the continuum field 
representation and the point particle representation are not equivalent.
The two systems obviously have different degrees of freedom and differ on the 
microscopic level, the crucial issue is whether or not they agree 
macroscopically as regards the formation of caustics. An Appendix (section 6) 
studies both systems for a very simple, spherically symmetric and static 
geometry, showing that the 4-velocity $u_{\mu}(t,r) = \partial_{\mu} 
\phi(t,r)$ does not form caustics, whereas caustics do occur for a collection 
of freely falling particles. Further studies on this problem are underway
for the field theory representation. Note however, that caustic formation is 
a UV problem which can in principle be cured by a suitable UV completion, 
while the philosophy our work is rather to feature an IR modification of 
gravity.

\subsection{The Energy Density}

With $u_{\mu} = \partial_{\mu} \phi$, we can express the covariant 
derivative of $u_{\mu}$ as $D_{\nu} u_{\mu} = D_{\nu} D_{\mu} \phi
= D_{\mu} u_{\nu}$. Hence the divergence of the dark matter stress
tensor is,
\begin{eqnarray}
D^{\nu} (\rho u_{\mu} u_{\nu}) & \!\!\! = \!\!\! & \rho u^{\nu}
D_{\nu} u_{\mu} + u_{\mu} D_{\nu} (u^{\nu} \rho) \; , 
\label{stress1} \\
& \!\!\! = \!\!\! & \tfrac12 \rho \partial_{\mu} (g^{\alpha\beta}
\partial_{\alpha} \phi \partial_{\beta} \phi) + 
\tfrac{\partial_{\mu} \phi}{\sqrt{-g}} \, \partial_{\alpha} \,
(\sqrt{-g} \, g^{\alpha\beta} \partial_{\beta} \rho) \; . 
\label{stress2}
\end{eqnarray}
Combining the scalar equation (\ref{phieqn}) with the conservation 
of the dark matter stress tensor implies a first order equation for 
the energy density $\rho(t,\vec{x})$,
\begin{equation}
\partial_{\mu} (\sqrt{-g} \, u^{\mu} \rho) = 0 \qquad , \qquad 
\rho(0,\vec{x}) = \tfrac{\rho_0}{\sqrt{{\rm det}[g_{ij}(0,\vec{x})]}}
\qquad , \qquad \rho_0 = \tfrac{45 a_0^2}{16 \pi G} \; . \label{rhoeqn}
\end{equation}
The initial condition was chosen to provide a nearly homogeneous 
energy density, with small perturbations driven by the primordial 
density perturbations, and the correct overall magnitude to replace 
cold dark matter,
\begin{equation}
\rho_0 \simeq \tfrac56 \times \tfrac{3}{10} \times \rho_{\rm crit}
= \tfrac{3 c^2 H_0^2}{32 \pi G} \simeq \tfrac{45 a_0^2}{16 \pi G} \; .
\label{rhoDM}
\end{equation}
The final equality exploits the numerical coincidence between $c H_0
\simeq 6.6 \times 10^{-10}~{\rm m/s}^2$ and Milgrom's constant $a_0 
\simeq 1.2 \times 10^{-10} {\rm m/s}^2$, which characterizes the 
transition from Newtonian gravity in MOND (MOdified Newtonian Dynamics) 
\cite{Milgrom:1983ca,Milgrom:1983pn,Milgrom:1983zz}. This nonlocal, 
modified gravity model automatically recovers the cosmological successes 
of dark matter without the problematic fundamental particle nature, 
which has so far eluded detection \cite{PandaX-II:2017hlx,XENON:2018voc,
PandaX-4T:2021bab,ADMX:2021nhd,LZ:2022lsv,PerezAdan:2023rsl,XENON:2023cxc}. 

Substituting the ADM form for the metric in (\ref{phieqn}) and 
employing the scalar equation (\ref{phieqn2}) gives,
\begin{equation}
\partial_{t} \Bigl[ \rho \sqrt{\gamma} \sqrt{1 + \gamma^{jk} \partial_j
\phi \partial_k \phi} \, \Bigr] = \partial_i \Bigl[N \rho \sqrt{\gamma}
\gamma^{ij} \partial_j \phi - N^i \rho \sqrt{\gamma} \sqrt{1 + \gamma^{jk}
\partial_j \phi \partial_k \phi} \, \Bigr] \; . \label{rhoeqn2}
\end{equation}
This is obviously a well-posed equation, which determines the energy density
$\rho[g](t,\vec{x})$ as a unique, nonlocal functional of the metric obeying
the initial condition $\rho(0,\vec{x}) = \rho_0/\sqrt{\gamma(0,\vec{x})}$.

Equations (\ref{phieqn}) and (\ref{rhoeqn}) determine the dark matter stress
tensor $T_{\mu\nu}[g] = \rho[g] \partial_{\mu} \phi[g] \partial_{\nu} 
\phi[g]$ as a nonlocal functional of the metric. This model necessarily
reproduces all of the cosmological successes of dark matter, including the
observed spectrum of anisotropies in the cosmic microwave radiation, baryon
acoustic oscillations and linearized structure formation. However, it is
worth noting that these successes derive completely from the dependence of
$\phi[g]$ and $\rho[g]$ on linearized perturbations of the metric about the 
cosmological background. Not only is the full, nonlinear model unnecessary,
it is not even particularly desirable. Because the model defined by
(\ref{phieqn}) and (\ref{rhoeqn}) is just dark matter, expressed as a 
nonlocal functional of the metric, it would suffer from the usual problems
of explaining the many observed regularities of gravitationally bound 
systems \cite{Brada:1998mr,Brada:1998mi,Milgrom:2009bi,McGaugh:2013zqa,
Milgrom:2012xw,Lelli:2016uea,Milgrom:2016ogb,Lelli:2016cui,McGaugh:2020ppt,
Banik:2021woo} such as the Baryonic Tully-Fisher Relation (BTFR) 
\cite{McGaugh:2000sr}. The presence of Milgrom's constant in expression 
(\ref{rhoeqn}) suggests that an attempt be made to extend the model so that 
it degenerates to MOND in the static limit. 

\subsection{A Lagrangian Formalism}

Our derivation relied on three assumptions:
\begin{enumerate}
\item{The stress tensor is separately conserved;}
\item{The 4-velocity is the gradient of a scalar $\phi$; and}
\item{The stress tensor agrees with the initial distribution of dark matter.}
\end{enumerate}
It is useful to note that the first two assumptions follow from a simple
Lagrangian in which one regards $\phi$ and $\rho$ as independent, local fields,
rather than as nonlocal functionals of the metric,
\begin{equation}
\mathcal{L} = -\tfrac12 \rho [\partial_{\mu} \phi \partial_{\nu} \phi
+ 1] \sqrt{-g} \; . \label{mimetic}
\end{equation}
Variation with respect to $\rho$ and $\phi$ gives the $\phi$ and $\rho$
equations,
\begin{equation}
\tfrac{\delta S}{\delta \rho} = -\tfrac12 [\partial_{\mu} \phi \partial_{\nu} 
\phi + 1] \sqrt{-g} = 0 \qquad , \qquad \tfrac{\delta S}{\delta \phi} = 
\partial_{\mu} [\sqrt{-g} \, g^{\mu\nu} \partial_{\nu} \phi \, \rho] = 0 \; .
\label{mimeticeqns}
\end{equation}
The associated stress tensor is,
\begin{equation}
-\tfrac{2}{\sqrt{-g}} \tfrac{\delta S}{\delta g^{\mu\nu}} = \rho \partial_{\mu}
\phi \partial_{\nu} \phi - \tfrac12 g_{\mu\nu} \rho [\partial_{\alpha} \phi
\partial_{\beta} \phi + 1] = \rho u_{\mu} u_{\nu} \; . \label{mimeticstress}
\end{equation}
The final equality results from the equation of motion (\ref{mimeticeqns}) and
the identification $u_{\mu} = \partial_{\mu} \phi$.  

Expression (\ref{mimetic}) is the Lagrangian for ``mimetic gravity'' 
\cite{Chamseddine:2013kea,Chamseddine:2014vna}, variations of which have been
much studied \cite{Sebastiani:2016ras,Domenech:2025qny}. It differs from our 
model in that the mimetic fields $\phi$ and $\rho$ are independent, with 
arbitrary initial value data, whereas ours fields are unique nonlocal 
functionals of the metric defined by expressions (\ref{phieqn}) and 
(\ref{rhoeqn}). However, it is important to note that the stress tensor 
(\ref{mimeticstress}) does not depend upon this distinction.
Note also that the simple mimetic Lagrangian (\ref{mimetic}) permits a 
detailed analysis of perturbations that rules out instabilities, ghosts and 
runaway behavior \cite{Barvinsky:2013mea}.

\section{The Gravitationally Bound Regime}

The purpose of this section is to review the successful model for 
gravitationally bound systems \cite{Deffayet:2011sk,Deffayet:2014lba}.
We begin by exploiting the Baryonic Tully-Fisher Relation to infer the
$g_{00}$ equation for gravitationally bound systems whose acceleration
is comparable to or less than Milgrom's constant $a_0 \simeq 1.2 \times 
10^{-10}~{\rm m/s}^2$, assuming there is no dark matter. Gravitational
lensing implies that the remaining Einstein equations are unmodified.
These two points are made in the context of a geometry which is static 
and spherically symmetric,
\begin{equation}
ds^2 = -(1 + 2 \Psi) c^2 dt^2 + (1 + 2 \Phi) d\vec{x} \cdot d\vec{x} 
\; . \label{static}
\end{equation}
The section closes by giving a Lagrangian whose addition to that of
general relativity would produce the phenomenologically correct
field equations for gravitationally bound systems.

\subsection{Requirements of the BTFR}

Let $\varrho(r)$ denote the baryonic mass density, which we
emphasize is completely different from the CDM energy density $\rho(x)$
discussed in the previous section. The baryonic mass enclosed at radius 
$r$ is,
\begin{equation}
M(r) = 4\pi \!\! \int_0^r \!\!\! ds \, s^2 \varrho(s) \; . 
\label{Mincl}
\end{equation}
An object undergoing circular motion at radius $r$ with velocity
$v(r)$ has centripetal acceleration,
\begin{equation}
\tfrac{v^2(r)}{r} = c^2 \Psi'(r) \qquad \Longrightarrow \qquad
v^2(r) = c^2 r \Psi'(r) \; . \label{vsquared}
\end{equation}
According to the Baryonic Tully-Fisher Relation \cite{McGaugh:2000sr},
\begin{equation}
v^4(r) = [c^2 r \Psi'(r)]^2 = a_0 G M(r) \; . \label{BTFR1}
\end{equation}
Differentiating uncovers the baryonic mass density $\varrho(r)$,
\begin{equation}
\tfrac{\partial}{\partial r} [c^2 r \Psi'(r)]^2 = a_0 G \times
4\pi r^2 \varrho(r) \; . \label{BTFR2}
\end{equation}
Rearranging the factors results in an equation for $\Psi(r)$ that
must pertain in order to reproduce the Baryonic Tully-Fisher Relation
without dark matter,
\begin{equation}
\tfrac{2 c^2}{a_0 r^2} \tfrac{\partial}{\partial r} [r \Psi'(r)]^2
= \tfrac{8\pi G}{c^2} \, \varrho(r) \; . \label{BTFR3}
\end{equation}
From the static geometry (\ref{static}) we see that (\ref{BTFR3}) 
should represent the $g_{00}$ equation of gravity.

\subsection{Requirements of Weak Lensing}

Equation (\ref{BTFR3}) is not at all what one gets from general
relativity. To lowest order in the two potentials of the static
geometry (\ref{static}), the nontrivial Einstein equations are,
\begin{eqnarray}
G_{00} & = & -2 \nabla^2 \Phi = \tfrac{8 \pi G}{c^2} \varrho \; , 
\label{G00} \\
G_{ij} & = & (\delta_{ij} \nabla^2 - \partial_i \partial_j) 
(\Psi + \Phi) = 0 \; . \label{Gij}
\end{eqnarray}
The spatial equations imply $\Phi = -\Psi$, which is consistent
with weak lensing provided that $\Psi$ obeys (\ref{BTFR3}), rather than
(\ref{G00}). To recover the desired equation (\ref{BTFR3}), one must add
a contribution (\ref{generalform}) to the gravitational Lagrangian whose
variation with respect to $\Phi$ does not disturb (\ref{Gij}) and whose
variation with respect to $\Psi$ cancels the linear term in (\ref{G00}),
which could be written in terms of $\Psi = -\Phi$. The desired form is
\cite{Deffayet:2011sk},
\begin{equation}
\Delta \mathcal{L} = \tfrac{c^4}{16 \pi G} \Bigl[ 2 {\Psi'}^2 -
\tfrac{4 c^2}{3 a_0} {\Psi'}^3 + \dots \Bigr] \sqrt{-g} \; .
\label{staticM}
\end{equation}

\subsection{An Invariant Lagrangian Formulation}

We seek an invariant Lagrangian whose specialization to the static
geometry reduces to (\ref{staticM}). The first task is finding a 
nonlocal invariant that interpolates the Newtonian potential $\Psi$. 
There are many ways of accomplishing this. A particularly simple 
choice is motivated by the nonzero components of the Ricci tensor 
to lowest order in the static potentials $\Psi$ and $\Phi$,
\begin{eqnarray}
R_{00} & \!\!\! = \!\!\! & \nabla^2 \Psi + \dots \; , \label{R00} \\
R_{ij} & \!\!\! = \!\!\! & -\delta_{ij} \nabla^2 \Phi - \partial_i
\partial_j (\Psi + \Phi) + \dots \; . \qquad \label{Rij}
\end{eqnarray}
One can isolate the $00$ component by contracting with the timelike
4-velocity, $u_{\mu}(x) = \partial_{\mu} \phi[g](x)$, where the scalar 
$\phi[g](x)$ was defined in equation (\ref{phieqn}). Then the factor
of $\nabla^2$ can be stripped off, when acting on time independent
functions in the static geometry (\ref{static}), using the inverse 
scalar d'Alembertian \cite{Deffayet:2011sk},
\begin{equation}
\square \equiv \tfrac{1}{\sqrt{-g}} \partial_{\mu} (\sqrt{-g} \, 
g^{\mu\nu} \partial_{\nu}) \longrightarrow \nabla^2 \qquad , \qquad
\tfrac{1}{\square} (R_{\alpha\beta} u^{\alpha} u^{\beta}) 
\longrightarrow \Psi \; .
\label{Psiinterp}
\end{equation}
Of course the Ricci tensor is not static for general metrics. The 
inverse scalar d'Alembertian becomes unique if we define it and its 
first derivative to vanish on the same $t=0$ initial value surface 
that was employed in equations (\ref{phieqn}) and (\ref{rhoeqn}). 

One can achieve an invariant realization of (\ref{staticM}) using an
algebraic function of the nonlocal invariant $Z[g](x)$ 
\cite{Deffayet:2011sk},
\begin{equation}
Z[g] \equiv \tfrac{4 c^4}{a_0^2} g^{\mu\nu} \partial_{\mu} \Bigl[
\tfrac1{\square} R_{\alpha\beta} u^{\alpha} u^{\beta} \Bigr] \partial_{\nu}
\Bigl[\tfrac1{\square} R_{\rho\sigma} u^{\rho} u^{\sigma} \Bigr]
\longrightarrow \tfrac{4 c^4}{a_0^2} \vec{\nabla} \Psi \! \cdot\!
\vec{\nabla} \Psi \; . \label{Zdef}
\end{equation}
The appropriate realization is,
\begin{equation}
\Delta \mathcal{L} \longrightarrow \tfrac{a_0^2}{16 \pi G} \!\times\!
f\Bigl( Z[g]\Bigr) \!\times\! \sqrt{-g} \; , \label{invL}
\end{equation}
where the small $Z > 0$ expansion of $f(Z)$ is,
\begin{equation}
f(Z) = \tfrac12 Z - \tfrac16 Z^{\frac32} + O(Z^2) \label{fexp}
\end{equation}
MOND phenomenology requires $f(Z)$ to be strongly suppressed when the 
acceleration $c^2 \Psi'(r)$ is greater than $a_0$. It turns out that we 
also want it to be suppressed in the cosmological regime for which 
$Z[g]$ is large and negative (see the appendix, which is section 7).
A simple function which accomplishes all three 
things is \cite{Deffayet:2011sk},
\begin{equation}
f(Z) = \tfrac12 Z \exp\Bigl[-\tfrac13 \sqrt{\vert Z\vert} \, \Bigr] \; .
\label{fdef}
\end{equation}

\section{Synthesis}

The purpose of this section is to exhibit a single MOND addition to the
Lagrangian of general relativity that interpolates between the 
successful cosmological model described in section 2 and the successful 
gravitationally bound model described in section 3. The general form is, 
\begin{equation}
\mathcal{L}_{\rm MOND} = -\tfrac{a_0^2}{16 \pi G} \times \mathcal{M}[g] 
\times \sqrt{-g} \; , \label{generalform}
\end{equation}
The gravitationally bound model (\ref{invL}) is already in this form so 
we begin by showing how the cosmological model (\ref{mimetic}) can be
written in the same way. We then exhibit a single equation for the nonlocal
functional $\mathcal{M}[g]$ which recovers the cosmological solution for
geometries whose time dependence dominates over space dependence, and 
reduces to the gravitationally bound solution when the reverse is true.
The section closes with a discussion of potential problems.

\subsection{The Complete Model} 

To express the cosmological Lagrangian (\ref{mimetic}) in the general form
we add a surface term,
\begin{eqnarray}
\lefteqn{-\tfrac12 \rho [\partial_{\mu} \phi \partial_{\nu} \phi + 1] \sqrt{-g}
+ \partial_{\mu} [\sqrt{-g} \, g^{\mu\nu} \partial_{\nu} \phi \, \rho \phi] 
= -\rho \sqrt{-g} } \nonumber \\
& & \hspace{3cm} + \tfrac12 \rho [\partial_{\mu} \phi \partial_{\nu} \phi 
g^{\mu\nu} + 1] \sqrt{-g} + \phi \partial_{\mu} [\sqrt{-g} \, g^{\mu\nu} 
\partial_{\nu} \phi \, \rho] \; . \label{newL} \qquad 
\end{eqnarray}
The two terms on the second line of (\ref{newL}) vanish with the equations of
motion (\ref{mimeticeqns}), which demonstrates that regarding (\ref{generalform})
as a nonlocal addition to the gravitational Lagrangian will reproduce the
cosmological model for $\mathcal{M}[g] = 45 \rho[g]/\rho_0$, with $\rho[g]$
defined by equations (\ref{phieqn}) and (\ref{rhoeqn}). Recall from
(\ref{rhoeqn}) that $\rho_0 \equiv 45 a_0^2/16\pi G$.

The single equation that defines $\mathcal{M}[g]$ generally is, 
\begin{equation}
\partial_{\mu} \Bigl[ \sqrt{-g} \, u^{\mu} \mathcal{M}\Bigr] = -
\partial_{\mu} \Bigl[ \sqrt{-g} \, u^{\mu} f\Bigl(Z[g]\Bigr) \Bigr] 
\qquad , \qquad \mathcal{M}(0,\vec{x}) = 
\tfrac{45}{\sqrt{{\rm det}[g_{ij}(0,\vec{x})]}} \; . \label{Meqn}
\end{equation}
Here $u_{\mu}[g] = \partial_{\mu} \phi[g]$ is defined by equation 
(\ref{phieqn}), $Z[g]$ is defined by equation (\ref{Zdef}), and 
expression (\ref{fdef}) gives the function $f(Z)$. In the cosmological
regime $Z[g]$ is large and negative, which makes $f(Z)$ infinitesimal,
so one recovers the successful cosmological model. Space dependence 
dominates inside gravitationally bound systems. Because the spatial 
components of the vector field $u^{\mu}[g]$ are nonzero, equation 
(\ref{Meqn}) pushes $\mathcal{M}[g]$ away from the homogeneous
solution towards $\mathcal{M}[g] \simeq -f(Z[g])$, which recovers
the successful model for gravitationally bound systems.

\subsection{Doing the Math}

Several issues deserve comment. The first of these concerns the crucial
mechanics of the transition from dominance of the positive homogeneous 
solution of equation (\ref{Meqn}) for cosmology to the negative, and
numerically larger, inhomogeneous solution of $\mathcal{M}[g] \simeq 
-f(Z[g])$ for a gravitationally bound system. Although a definitive 
analysis in a realistic setting is obviously a formidable numerical 
undertaking, a qualitative understanding of the process can probably 
be gained by working it out for a fictitious geometry that interpolates 
between cosmology and gravitational binding. We have in mind a 
spherically symmetric metric of the general form,
\begin{equation}
ds^2 = -B(t,r) dt^2 + A(t,r) d\vec{x} \!\cdot\! d\vec{x} \; , 
\label{fake}
\end{equation}
where the functions $A(t,r)$ and $B(t,r)$ smoothly interpolate from 
homogeneous expansion ($A \rightarrow a^2(t)$ and $B \rightarrow$ 
constant) for large $r$ to the static geometry (\ref{static}) for small 
$r$. The project becomes a tractable, 2-dimensional problem if one 
simply {\it invents} the functions $A(t,r)$ and $B(t,r)$, without 
requiring them to satisfy any particular gravitational field equations. 
The analysis would consist of solving equation (\ref{phieqn2}) for 
$\phi(t,r)$ in this geometry, and then equations (\ref{Zdef}) for 
$Z(t,r)$ and (\ref{Meqn}) for $\mathcal{M}(t,r)$. It will probably
require some trial and error to find simple and compelling candidates
for $A(t,r)$ and $B(t,r)$, but such a study would make a nice followup 
to the current work, and should clarify how the transition takes 
place.

A second general issue concerns stability. The perturbations of mimetic
gravity were studied by Barvinsky \cite{Barvinsky:2013mea}, with the 
result that there are no instabilities, ghosts, or runaway behaviors.
A similar analysis of perturbations for the MOND model of section 4 has
been conducted with the same result \cite{Tan:2018bfp}. 

A final issue is how the modification affects the propagation of tensor
modes. This is strongly constrained by multimessenger observations
\cite{Goldstein:2017mmi,LIGOScientific:2017vwq,LIGOScientific:2017ync,
LIGOScientific:2017zic}, which falsify some realizations of MOND 
\cite{Boran:2017rdn}. However, the modification we propose makes no
change in the propagation of tensor modes. For the cosmological model of
section 3 this follows from the analysis Barvinsky has made of our
version of mimetic gravity \cite{Barvinsky:2013mea}. The same conclusion
follows for the MOND realization of section 4 from the simple fact that
its source, $Z[g]$, contains two factors of the Ricci tensor, which 
vanishes for gravitational radiation.

\section{Conclusions}

We have presented a fully relativistic, modified gravity model that
interpolates between MOND, inside gravitationally bound systems, and
a pressureless, perfect fluid for cosmology, which mimics dark matter. 
Our model consists of a nonlocal addition (\ref{generalform}) to the 
Lagrangian of gravity, where the functional $M[g](x)$ obeys equation 
(\ref{Meqn}) and the timelike 4-velocity field $u_{\mu}[g] = 
\partial_{\mu} \phi[g]$, is defined by (\ref{phieqn}). The functional 
$Z[g]$ is defined by equation (\ref{Zdef}), with null initial value 
data, and the function $f(Z)$ is given in equation (\ref{fdef}).

Section 2 describes the cosmological regime, which reproduces the successes 
of dark matter, including acoustic oscillations during decoupling, bar\-y\-on 
acoustic oscillations, and linearized structure formation. The model cannot
be distinguished from dark matter because it was constructed by exploiting
conservation to express the dark matter stress tensor as a nonlocal
functional of the metric. This automatically reproduces the gravitational
signatures of dark matter without the problematic fundamental fields, which 
have stubbornly eluded detection \cite{PandaX-II:2017hlx,XENON:2018voc,
PandaX-4T:2021bab,ADMX:2021nhd,LZ:2022lsv,PerezAdan:2023rsl,XENON:2023cxc}. 

Similarly, in section 3 we described a relativistic realization of MOND for
the gravitationally bound systems. Like the cosmological model of section 2,
this has to work because it was constructed to enforce weak lensing and the 
equation (\ref{BTFR3}) required by the Baryonic Tully-Fisher Relation. 
Although the models of section 2 and section 3 had to work in their 
respective regimes, it is significant that there is a smooth interpolation 
between them. That was described in section 4.

Cosmology used to be held up as an impossible goal for a relativistic 
realization of MOND \cite{Pardo:2020epc}, however, we can now see that {\bf
this is false.} In addition to the model described in this paper, there are 
two local models, involving vector and scalar fields in addition to the 
metric, that succeed in describing cosmology \cite{Blanchet:2008fj,
Blanchet:2009zu,Blanchet:2012ey,Skordis:2020eui,Skordis:2021mry,
Verwayen:2023sds,Bataki:2023uuy,Durakovic:2023out,Blanchet:2024mvy}.

This paper suggests a number of follow-up projects. The first of these is
to explore the transitions between the three regimes:
\begin{enumerate} 
\item{{\bf Cosmology:} characterized by $Z[g] < 0$;}
\item{{\bf Deep MOND:} characterized by $0 < Z[g] \ltwid 1$; and}
\item{{\bf Newtonian:} characterized by $1 \ll Z[g]$.}
\end{enumerate}
The model was constructed to agree with data from deep inside each 
regime, but it may make distinctive and testable predictions for systems
between two regimes. Of particular importance is the issue
of structure formation beyond the linear regime, which concerns the 
transition from cosmology to gravitationally bound systems. Because the
interpolation mechanism may become relevant at intermediate scales, it
is not clear whether nonlinear clustering and lensing statistics remain
consistent with observations. A related project is to explore different 
possibilities for the interpolation function $f(Z)$ and for invariantly
realizing the (square of the) Newtonian acceleration in units of $a_0$,
$Z[g]$. The choices made in expressions (\ref{fdef}) and (\ref{Zdef}) 
seem adequate but it would be desirable to study what other possibilities
exist.

Two transition regions of great interest are the cores of galactic 
clusters \cite{Aguirre:2001fj} and colliding clusters \cite{Clowe:2006eq}.
These are traditionally regarded as problematic for MOND but they may
instead be the result of competition between the cosmological regime and
the deep MOND regime. That is, the homogeneous solution of (\ref{Meqn})
--- which describes cosmology --- may be comparable, for these systems,
to the inhomogeneous solution, which describes gravitationally bound
systems far from  the Hubble flow. Given the importance 
of cluster observations to the dark matter problem, a more detailed 
analysis is called for. In section 4.2 we have indicated a preliminary,
analytic followup project that should clarify, in rough terms, the 
transition of $\mathcal{M}[g]$ from the positive, homogeneous solution 
of cosmology to the negative, and numerically much larger, inhomogeneous 
solution of a gravitationally bound system. If this analysis indicates 
that galactic clusters fall in the regime of ``competition'' between the 
two solutions, it would serve as a powerful motivation for a realistic 
and detailed numerical study.

Another project is to explore the ``External Field Effect'' 
\cite{Milgrom:2012xw}, whereby the gravitational fields produced by
distant masses can affect whether or not a local system is in the 
MOND or Newtonian regimes. Without a relativistic extension of MOND
this could only be guessed at, but it can be addressed in detail 
within the context of a model such as we have presented. The key
issue is how distant masses affect the nonlocal functional $Z[g](x)$.

Finally, there is the need to derive $\mathcal{M}[g]$ from a 
nonperturbative resummation of loops of inflationary gravitons. The
effects of matter loops on de Sitter background have already been
subjected to such a resummation \cite{Woodard:2025smz}. What is 
needed is to extend the technique to general backgrounds, which has
already been done for nonlinear sigma models \cite{Kasdagli:2023nzj,
Woodard:2023cqi}, and to generalize it to graviton loops. Although
graviton loops are challenging, recent progress gives cause for
optimism \cite{Miao:2024shs,Miao:2025gzm,Miao:2025bmd}. It hardly
needs to be said that success would not only produce a definite
$\mathcal{M}[g]$, but also confirm MOND and falsify the dark matter
hypothesis.

\section{Appendix: Inequivalent Representations}

The purpose of this section is to demonstrate that the representation
of pressureless dark matter in terms of continuum fields $\phi(x)$
and $\rho(x)$, obeying equations (\ref{phieqn2}) and (\ref{rhoeqn}),
is not equivalent to the representation as an collection of freely
falling, point particles. This is relevant to the crucial issue of 
caustics, which certainly form as particles fall into gravitational
potentials, but do not seem possible for a 4-velocity $u_{\mu}(x) =
\partial_{\mu} \phi(x)$ in a smooth geometry. Because a single
counter-example suffices to prove the point, we consider the two 
systems in a static, spherically symmetric ADM geometry (\ref{ADM})
with,
\begin{equation}
0 < N(r) < 1 \qquad , \qquad N^i = 0 \qquad \gamma^{ij} = \delta^{ij} 
\; . \label{counterex1}
\end{equation}
We further assume $N(r)$ falls monotonically from $N(\infty) = 1$ to 
$N(0) = \frac12$, with $N'(0) = 0$. An example would be,
\begin{equation}
N(r) = 1 - \tfrac{R}{2 \sqrt{R^2 + r^2}} \; . \label{counterex2}
\end{equation}
However, it is not necessary to commit to (\ref{counterex2}). 

The scalar depends only on $t$ and $r$ and obeys the equation,
\begin{equation}
\dot{\phi}(t,r) = N(r) \sqrt{1 + [\phi'(t,r)]^2} \qquad , \qquad
\phi(0,r) = 0 \; . \label{counterex3}
\end{equation}
The only possibility for a caustic to form at r=0 would be if 
$\phi(t,r)$ had a cusp there. However, it is easy to see that 
$\phi'(t,0) = 0$ for all time. First, note that $\phi'(0,r) = 0$. 
Now differentiate equation (\ref{counterex3}) with respect to $r$,
\begin{equation}
\dot{\phi}'(t,r) = N'(r) \sqrt{1 + [\phi'(t,r)]^2} + \tfrac{N(r)
\phi'(t,r) \phi''(t,r)}{\sqrt{1 + [\phi'(t,r)]^2}} \; . 
\label{counterex4}
\end{equation}
Because all spatial derivatives of $\phi(0,r)$ vanish, we see that
$\dot{\phi}'(0,0) = 0$. Now differentiate (\ref{counterex4}) with
respect to time to conclude that $\ddot{\phi}'(0,0) = 0$. Because
each additional time derivative involves only lower time derivatives,
recursion implies that all time derivatives of $\phi'(t,r)$ vanish
at $(t,r) = (0,0)$.

On the other hand, it is simple to show that freely falling particle
geodesics oscillate through $r=0$, which leads to a caustic. To see
this note that the only nonvanishing components of the affine 
connection are,
\begin{equation}
\Gamma^0_{~0r} = \Gamma^0_{~r0} = \tfrac{N'}{N} \qquad , \qquad
\Gamma^r_{~00} = N N' \; . \label{counterex6}
\end{equation}
Geodesics are radial in this model, $\chi^{\mu}(\tau) = 
(t(\tau),r(\tau))$. An overdot denotes differentiation with respect 
to $\tau$, and the initial values that agree with the continuum
model are,
\begin{equation}
\dot{t}(0) \equiv \dot{t}_0 = \tfrac1{N(r_0)} \qquad , \qquad 
\dot{r}(0) \equiv \dot{r}_0 = 0 \; . \label{counterex7}
\end{equation}
The geodesic equation for $t(\tau)$ implies,
\begin{equation}
0 = \ddot{t} \!+\! \Gamma^0_{~\alpha\beta} \dot{\chi}^{\alpha} 
\dot{\chi}^{\beta} = \ddot{t} \!+\! 2 \tfrac{N'}{N} \dot{r} \dot{t} 
= \ddot{t} \!+\! \dot{t} \!\times\! \tfrac{d}{d\tau} \ln\Bigl[ 
N^2\Bigl(r(\tau)\Bigr)\Bigr] \Longrightarrow \dot{t}(\tau) = 
\tfrac{N(r_0)}{N^2(r(\tau))} \; . \label{counterex8}
\end{equation}
At this point the timelike condition fixes $\dot{r}(\tau)$,
\begin{equation}
g_{\mu\nu} \dot{\chi}^{\mu} \dot{\chi}^{\nu} = -1 \qquad 
\Longrightarrow \qquad \dot{r}(\tau) = \pm \sqrt{ \tfrac{N^2(r_0)}{
N^2(r(\tau))} - 1} \; . \label{counterex9}
\end{equation}
We see that each particle oscillates around the origin with
constant amplitude, causing there to be a caustic at $r=0$.

\section{Appendix: $Z[g](x)$ for Cosmology}

The purpose of this appendix is to work out the functional $Z[g](x)$,
defined by expression (\ref{Zdef}), for the special case of a 
cosmological geometry characterized by scale factor $a(t)$,
\begin{equation}
ds^2 = -c^2 dt^2 + a^2(t) d\vec{x} \!\cdot\! d\vec{x} \qquad 
\Longrightarrow \qquad H(t) \equiv \tfrac{\dot{a}}{a} \; . \label{FLRW}
\end{equation}
For this case the timelike 4-velocity is $u_{\mu} = \delta^0_{~\mu}$
and the various building blocks of $Z[g]$ degenerate to,
\begin{eqnarray}
R_{\alpha\beta} u^{\alpha} u^{\beta} & \!\!\! \longrightarrow \!\!\! &
-\tfrac{3}{c^2} (H^2 + \dot{H}) \; , \label{degen1} \\
\square f(t) & \!\!\! \longrightarrow \!\!\! & -\tfrac1{c^2 a^3} 
\partial_t (a^3 \dot{f}) \; , \label{degen2} \\
g^{\mu\nu} \partial_{\mu} f(t) \partial_{\nu} f(t) & \!\!\! 
\longrightarrow \!\!\! & -\tfrac1{c^2} \dot{f}^2 \; . \label{degen3}
\end{eqnarray}
Substituting (\ref{degen1}-\ref{degen3}) in the definition (\ref{Zdef})
of $Z[g]$ gives,
\begin{equation}
-Z[g] \longrightarrow \Bigl[ \tfrac{6 c}{a_0 a^3(t)} \int_0^{t} \!\! dt'
\, a^3(t') [H^2(t') + \dot{H}(t')] \Bigr]^2 \; . \label{cosmoZ}
\end{equation}

Expression (\ref{cosmoZ}) shows that $Z[g]$ is negative for cosmology.
To see that it is large at high redshift it is useful to convert from
comoving time to redshift and specialize to the $\Lambda$CDM model,
\begin{eqnarray}
1 + z & \!\!\! \equiv \!\!\! & \tfrac{a_0}{a(t)} \qquad \Longrightarrow
\qquad \tfrac{dz}{1 + z} = - H dt \; , \label{redshift} \\
H(z) & \!\!\! \equiv \!\!\! & H_0 \sqrt{ \Omega_{\rm r} (1 + z)^4
+ \Omega_{\rm m} (1 + z)^3 + \Omega_{\Lambda} } \; . \label{LCDM}
\end{eqnarray}
The parameters $\Omega{\rm r}$, $\Omega_{\rm m}$ and $\Omega_{\Lambda}$
are approximately \cite{Planck:2018vyg},
\begin{equation}
\Omega_{\rm r} \simeq 10^{-4} \quad , \quad \Omega_{\rm m} \simeq 
\tfrac{3}{10} \quad , \quad \Omega_{\Lambda} \simeq \tfrac{7}{10} \; .
\label{Onmegas}
\end{equation}
Substituting (\ref{redshift}-\ref{LCDM}) in (\ref{cosmoZ}) gives,
\begin{equation}
\sqrt{-Z} \longrightarrow \tfrac{6 c H_0}{a_0} (1 + z)^3 \!\! 
\int_{z}^{\infty} \!\!\!\! \tfrac{dz'}{(1 + z')^4} \tfrac{\Omega_{\rm r} 
(1 + z')^4 + \frac12 \Omega_{\rm m} (1 + z')^3 - \Omega_{\Lambda}}{
\sqrt{ \Omega_{\rm r} (1 + z')^4 + \Omega_{\rm m} (1 + z')^3 + 
\Omega_{\Lambda} } } \; . \label{ZLCDM}
\end{equation}
The expansion for large redshift is straightforward \cite{Kim:2016nnd},
\begin{equation}
\sqrt{-Z} \longrightarrow \tfrac{6 c H_0}{a_0} \sqrt{\Omega_{\rm r}}
(1 + z)^2 \Bigl\{1 + O([\tfrac{1 + z_{\rm eq}}{1 + z}]^2) \Bigr\} \qquad
, \qquad z_{\rm eq} \equiv \tfrac{\Omega_{\rm m}}{\Omega_{\rm r}} \; .
\end{equation}
Note that $\frac{6 c H_0}{a_0} \sqrt{\Omega_{\rm r}} \simeq \frac13$.

A final point is that vacuum energy domination causes the integrand in 
expression (\ref{ZLCDM}) to become negative. This occurs quite late in 
cosmic history. Even later, the accumulated positive contribution from
early times is canceled to make $Z$ vanish. Detailed numerical study shows
that this happens at $z \simeq 0.0880$ \cite{Kim:2016nnd}.

\vskip .2cm

\centerline{\bf Acknowledgements}

We are grateful for conversations on the issue of caustics
with E. Babichev, V. F. Mukhanov, T. Prokopec and A. Vikman. This 
work was partially supported by NSF grant PHY-2207514 and by the 
Institute for Fundamental Theory at the University of Florida.

\end{document}